\documentclass[final,5p,twocolumn]{elsarticle}

\journal{System and Control Letters}

\usepackage{amsmath,amssymb}
\usepackage{amsthm}
\usepackage{color}
\usepackage{xcolor}
\usepackage{enumerate}
\usepackage{todo}
\usepackage{graphicx}
\usepackage{tikz,pgfplots}
\usepackage[ruled,vlined]{algorithm2e}
\usepackage{url}
\usetikzlibrary{intersections, backgrounds}
\usepgfplotslibrary{groupplots}
\usepackage{comment}

\newtheorem{lem}{Lemma}
\newtheorem{thm}{Theorem}
\newtheorem{rem}{Remark}

\makeatletter
\newcommand{\pushright}[1]{\ifmeasuring@#1\else\omit\hfill$\displaystyle#1$\fi\ignorespaces}
\newcommand{\pushleft}[1]{\ifmeasuring@#1\else\omit$\displaystyle#1$\hfill\fi\ignorespaces}
\makeatother

\usepackage{Notation}
\usepackage{mathtools}

\begin{document}
	
	\begin{frontmatter}	
		\title{Structure-Preserving Model Order Reduction for Index Two Port-Hamiltonian Descriptor Systems  \tnoteref{t1,t2}}
        \tnotetext[t1]{The research by P. Schwerdtner, M. Voigt, and T. Moser was supported by the German Research Foundation (DFG) within the projects 424221635 and 418612884 and that of V. Mehrmann was supported by the DFG through project B03 of SFB TRR 154.}
        \tnotetext[t2]{\textbf{CRediT author statement:} \textbf{Tim Moser, Paul Schwerdtner:} Conceptualization, Methodology, Software, Data Curation, Writing -- Original Draft, Visualization, \textbf{Volker Mehrmann:} Conceptualization, Writing -- Review \& Editing, Supervision, \textbf{Matthias Voigt:} Conceptualization, Writing -- Review \& Editing, Supervision, Funding Acquisition}
		\author[TUM]{Tim Moser\corref{COR}}
		\ead{tim.moser@tum.de}	
		\author[TUB]{Paul Schwerdtner\corref{COR}}
		\ead{schwerdt@math.tu-berlin.de}	
		\author[TUB]{Volker Mehrmann}
		\author[FUS]{Matthias Voigt}	
		
		\affiliation[TUM]{organization={Technical University of Munich, Chair of Automatic Control},
			addressline={Boltzmannstra{\ss}e 15}, 
			city={85748 Garching/Munich},
			country={Germany}}			
			
		\affiliation[TUB]{organization={TU Berlin, Institute of Mathematics},
			addressline={Stra{\ss}e des 17.~Juni 136},
			city={10623 Berlin},
			%postcode={10623}, 
			country={Germany}}

		\affiliation[FUS]{organization={UniDistance Suisse},
			addressline={Schinerstrasse 18}, 
			city={3900 Brig},
			country={Switzerland}}					
		\cortext[COR]{Authors contributed equally. Corresponding authors.}	
		
		\begin{abstract}
                  We present a new optimization-based structure-preserving model order reduction (MOR) method for port-Hamiltonian descriptor systems (pH-DAEs) with differentiation index two. Our method is based on a novel parameterization that allows us to represent any linear time-invariant pH-DAE with a minimal number of parameters, which makes it well-suited to model reduction. We propose two algorithms which directly optimize the parameters of a reduced model to approximate a given large-scale model with respect to either the $\hinf$ or the $\htwo$ norm. This approach has several benefits. Our parameterization ensures that the reduced model is again a pH-DAE system and enables a compact representation of the algebraic part of the large-scale model, which in projection-based methods often requires a more involved treatment. The direct optimization is entirely based on transfer function evaluations of the large-scale model and is therefore independent of the system matrices' structure. Numerical experiments are conducted to illustrate the high accuracy and small reduced model orders in comparison to other structure-preserving MOR methods.

		\end{abstract}
		
		\begin{keyword}
			port-Hamiltonian systems \sep structure-preserving model order reduction \sep descriptor systems
		\end{keyword}
	\end{frontmatter}
	
	\section{Introduction}%
	\label{sec:introduction}
	
	In this article, we generalize previously developed  optimization-based structure-preserving model order reduction (MOR) algorithms for port-Hamiltonian differential algebraic systems (pH-DAEs) of differentiation index one~\cite{MosSMV22} to pH-DAEs with index two, which  occur very frequently in applications, see, e.\,g., \cite{Beattie2018}.
To accommodate the hidden constraints that may occur in 
higher-index pH-DAEs, we adapt both the parameterization of the reduced-order model (ROM) and the optimization strategy compared to the case of index one pH-DAEs, as considered in~\cite{MosSMV22}.

As a system class, we consider linear constant coefficient pH-DAE systems of the form
\begin{equation}\label{eq:FOM}
  \begin{aligned}
    \pE \dot x(t) &= (\pJ-\pR)x(t) +(\pG-\pP)u(t), \\
    y(t) &= {(\pG+\pP)}^\T x(t) +(\pS-\pN)u(t),
  \end{aligned}
\end{equation}
where $\pE,\, \pJ,\, \pR \in \R^{\nx \times \nx}$, $\pG,\, \pP \in \R^{\nx\times \enu}$, $\pS,\, \pN \in \R^{\enu\times \enu}$ and the quadratic Hamiltonian is $\mathcal{H}(x) = \frac{1}{2}x^\T E x$. The following conditions are imposed on the system matrices:
\begin{enumerate}[(i)]
  \item The matrices $J$ and $N$ are skew-symmetric.
  \item The \emph{passivity matrix}
    \begin{align*}
      \label{eq:passivitymatrix}
      W := \mat{\pR & \pP \\ \pP^\T & \pS}
    \end{align*}
    and the \emph{energy matrix} $E$ are symmetric positive semi-definite.
\end{enumerate}

Note that we consider systems with $\pQ=I_n$ rather than the more general definition of pH-DAEs in \cite{MehU2022} governed by a quadratic Hamiltonian of the form $\mathcal{H}(x) = \frac{1}{2}x^\T Q^\T E x$ with $\pQ^\T \pE\geq 0$. However, this does not pose any real restriction, since it has been shown that every linear time-invariant pH-DAE may be reformulated to display this form; see~\cite{Mehl2021a} for a detailed discussion. 

In this paper, we treat systems that satisfy the following additional assumptions: (i) the pencil $sE - (J-R)$ is \emph{regular}, i.\,e.,  $\det(sE-(J-R))$ is not the zero polynomial; and (ii) all finite eigenvalues of $sE-(J-R)$ have a negative real part. 

PH-DAE systems of this type occur in energy-based modeling in many different physical domains, such as electrical systems~\cite{MehMOS18}, flow-problems~\cite{DomHLMMT21, HauMMMMRS19}, or mechanical multi-body systems~\cite[Example 12]{Beattie2018}. The  pH formulation is particularly useful where systems from different physical domains or of different scales and accuracy are coupled. Since the pH structure is preserved for parallel and negative feedback interconnections of pH systems and the index of pH-DAEs is at most two as shown in~\cite{Mehl2018Indexatmost2}, pH-DAEs can be safely coupled without increasing the index. In contrast, for general DAEs, the index may increase arbitrarily in such a coupling, and eigenvalues may move to the right-half plane, leading to an unstable system behavior. This emphasizes the usefulness of pH-DAEs when modeling subsystems in large networks of dynamical systems as in~\cite{DomHLMMT21,DaeMRS21,MehSS18}. 

When a pH-DAE describes a complex physical process or a complex geometry arises during modeling, the system typically has a high state-space dimension $\nx$, which renders the direct simulation or model-based controller design computationally expensive. In this case, MOR is used to compute reduced-order models (ROMs) that lead to similar output trajectories as their full-order model (FOM) counterparts when the same inputs are provided. The ROM can then be used in simulations and for control instead of the FOM\@.

The input-to-output behavior of a linear system in the frequency domain is described by its \emph{transfer function}. Under the assumption of regularity of the pencil $sE-(J-R)$, the transfer function of the system in~\eqref{eq:FOM} is given by
\begin{equation*}
  \pHtf(s) = {(\pG+\pP)}^\T {(s \pE-(\pJ-\pR))}^{-1}(\pG-\pP)+(\pS-\pN).
\end{equation*}
The typical use-case of pH systems in networks requires a structure-preserving MOR of pH-DAEs, i.\,e., the ROM must also be a pH system. Otherwise, the intuitive coupling to the rest of the network system is not possible at the ROM level. In particular, the ROM is supposed to be of the form
\begin{align*}
  \begin{split}
     \rE \rdx(t) &= (\rJ-\rR)\rx(t) +(\rG-\rP)u(t), \\
    \ry(t) &= {(\rG+\rP)}^\T \rx(t) +(\rS-\rN)u(t),
  \end{split}
\end{align*}
where the matrices $E_r,\, J_r,\, R_r \in \R^{r \times r}$, $G_r,\, P_r \in \R^{r\times \enu}$, $S_r,\, N_r \in \R^{\enu\times \enu}$ satisfy the same assumptions as the system in~\eqref{eq:FOM} with $r \ll \nx$ and for each admissible input function $u$ we have that $y \approx \ry $. The latter condition can be expressed as the requirement that the transfer function of the ROM should be a good approximation of the transfer function of the FOM\@.

The transfer function $H$ can be decomposed as
\begin{align*}
  \label{eq:introtfdecomp}
  \pHtf(s) = H_{\text{sp}}(s) + H_{\text{pol}}(s),
\end{align*}
where $H_{\text{sp}}$ is a \emph{strictly proper} rational matrix function and $H_{\text{pol}}$ is a matrix polynomial of degree $\nu-1$. Here, $\nu$ denotes the differentiation index of the uncontrolled DAE. The additional polynomial part in a FOM transfer function poses an extra challenge for MOR of DAEs because it must be matched \emph{exactly} in the ROM transfer function. Otherwise the error between FOM and ROM can become unbounded (see Section~\ref{sec:preliminaries} for error measures). Therefore, in projection-based MOR, the algebraic part of the DAE is typically treated separately, see~\cite{BeaGM21,Benner2017Overview,Gugercin2013DAEInterp,HauMM19,MehS05}. While for $\nu = 1$, the additional polynomial part is always constant, such that the FOM system can still be approximated with an ODE system with feedthrough term, for $\nu = 2$, the transfer function of a pH-DAE model  includes a linear term, which requires a reduced-order DAE model.

While the MOR of general higher-index DAEs has been extensively studied since the early 2000s (see~\cite{Stykel2004Gramian} for balancing-based methods and~\cite{Gugercin2013DAEInterp, Benner2006Partial} for interpolation-based methods), the structure-preserving MOR of pH-DAEs has still only partially been resolved. The methods in~\cite{BeaGM21,Hauschild2019} require a separate treatment of the algebraic part in order not to destroy the pH structure and often lead to an accuracy several orders of magnitude lower compared to non-structure-preserving MOR methods. Furthermore, the method in~\cite{BeaGM21} does not guarantee preservation of the pH structure when the FOM transfer function is improper and is only generally applicable to pH-DAEs where the hidden constraints are uncontrolled.

An alternative approach to pH structure-preserving MOR is passivity-preserving MOR, since the pH structure can often be recovered from a passive ROM~\cite{Beattie2018}. However, passivity-preserving methods such as positive-real balanced truncation (PRBT), as in~\cite{Reis2010}, often require an extra treatment of the algebraic equations which may impose an additional computational burden. Furthermore, transformations to the FOM matrices are needed that either require a priori knowledge of the given DAE structure or the construction of spectral projectors~\cite{MehS05}. 

This article describes an optimization-based MOR approach for structure-preserving MOR that
\begin{enumerate}[(i)]
  \item does not require transformations of the original system matrices and instead only requires evaluations of the FOM transfer function,
  \item addresses improper, proper, and strictly proper pH-DAEs of any type in one framework,
  \item leads to high fidelity pH-ROMs in the $\htwo$ and $\hinf$ norms with low model order, and 
  \item preserves any polynomial part in the FOM transfer function with the minimal possible number of states.
\end{enumerate}

The paper is organized as follows.
In Section~\ref{sec:preliminaries} we provide a short background on MOR and explain the properties and different types of pH-DAEs using a staircase form. In Section~\ref{sec:our_method} we present and explain our new method, which includes a parameterization of pH-DAEs and two optimization strategies implemented in Algorithms~\ref{alg:sobmor} and~\ref{alg:H2OPT}. Numerical experiments that illustrate the effectiveness of our approach are presented in Section~\ref{sec:numerical_examples}.

	\section{Preliminaries}%
	\label{sec:preliminaries}
	
	In this section we present some preliminary results for pH-DAEs that form the basis of the construction of the MOR algorithms.
We then summarize several MOR techniques for DAEs and their properties. 

\subsection{Staircase form}\label{sec:stc}
Our first result resembles a staircase form for pH-DAEs, which was derived for general linear pH-DAE systems without input and output in \cite{Achleitner2021} and for the system with input and output in \cite{BeaGM21}, see also \cite{MehU2022}.
The staircase form allows us to determine the differentiation index of a given pH-DAE but is
based on subsequent rank decisions which may be a challenging task in finite precision arithmetic.
Fortunately, the staircase form often emerges naturally from the modeling process~\cite{Beattie2018,Guducu2021} or can be constructed using only a few permutations of the given system matrices. We present the real-valued version of the form as derived in~\cite{Achleitner2021}. 

\begin{lem}\label{lem:staircase} \textup{\cite{Achleitner2021}}
Consider a regular pH-DAE system as in~\eqref{eq:FOM}. Then there exists an orthogonal  matrix $T\in\R^{\nx\times \nx}$ such that
\begin{align*}
	\sE &:= T\pE T^\T = \mat{\pE_{11} & \pE_{12} & 0 & 0 \\ \pE_{21} & \pE_{22} & 0 & 0 \\ 0 & 0 & 0 & 0 \\ 0 & 0 & 0 & 0}, \\
	\sJ &:= T\pJ T^\T = \mat{\pJ_{11} & \pJ_{12} & \pJ_{13} & \pJ_{14} \\  \pJ_{21} & \pJ_{22} & \pJ_{23} & 0 \\ \pJ_{31} & \pJ_{32} & \pJ_{33} & 0 \\ \pJ_{41} & 0 & 0 & 0}, \\
	\sR &:= T\pR T^\T = \mat{\pR_{11} & \pR_{12} & \pR_{13} & 0 \\  \pR_{21} & \pR_{22} &\pR_{23} & 0 \\ \pR_{31} & \pR_{32} & \pR_{33} & 0 \\ 0 & 0 & 0 & 0}, \\
	\sG &:= T\pG = \mat{\pG_1 \\\pG_2 \\ \pG_3 \\ \pG_4}, \sP := T\pP = \mat{\pP_1 \\\pP_2 \\ \pP_3 \\ 0}, \,\sS:=S,\,\sN:=N,
\end{align*}
where $\begin{bsmallmatrix}\pE_{11} & \pE_{12} \\ \pE_{21} & \pE_{22}\end{bsmallmatrix}$ is positive definite and the matrices $\pJ_{41}$ and $\pJ_{33}-\pR_{33}$ are invertible (if the blocks are nonempty).

The transformed system
\begin{equation}\label{eq:FOM_stair}
	\begin{aligned}
		\sE \dot{\sx}(t) &= (\sJ-\sR)\sx(t) +(\sG-\sP)u(t), \\
		´y(t) &= {(\sG+\sP)}^\T \sx(t) +(\sS-\sN)u(t),
	\end{aligned}
\end{equation}
is again a pH-DAE system with accordingly partitioned state vector $\sx(t) = {\left[\sx_1(t)^\T,\sx_2(t)^\T,\sx_3(t)^\T,\sx_4(t)^\T\right]}^\T$, where $\sx_i(t)\in\R^{\nx_i}, \nx_i\in \mathbb{N}_0$ for all $i=1,\ldots,4$.
The differentiation index $\ind$ of the uncontrolled system satisfies
\begin{equation*}
	\ind = \begin{cases}
		0 & \textit{if and only if } \nx_1 = \nx_4 = 0 \textit{ and } \nx_3 = 0, \\ 1 & \textit{if and only if } \nx_1=\nx_4=0 \textit{ and } \nx_3 > 0, \\ 2 & \textit{if and only if } \nx_1 = \nx_4 > 0.
	\end{cases} 
\end{equation*}
\end{lem}	

\subsection{Model order reduction of higher-index systems}%
\label{sub:model_order_reduction_of_higher_index_systems}
MOR for descriptor systems of possibly higher index have been proposed by many authors, see, e.\,g., \cite{Gugercin2013DAEInterp,MehS05,Stykel2004Gramian,BenW18,BanS14,Sorensen2005,breiten2021passivity}, to mention but a few. In such methods, the error between the transfer function of the FOM and the ROM are typically measured in norms defined in the Hardy spaces $\rhinf^{p \times m}$ and $\mathcal{RH}_2^{p \times m}$ of all proper (resp. strictly proper) real-rational $p \times m$ matrices without poles in the closed complex half-plane $\overline{\C^+} := \left\{ s \in \C\; |\; \operatorname{Re}(s) \ge 0 \right\}$. The typical norms are 
\begin{align*}
	\quad \left\| \pHtf \right\|_{\mathcal{H}_\infty} &:= \sup_{\omega \in \R} {\|\pHtf(\mathsf{i}\omega)\|}_2, \\
	\left\| \pHtf \right\|_{\mathcal{H}_2} &:= {\left(\frac{1}{2\pi}\int_{-\infty}^\infty {\|\pHtf(\mathsf{i}\omega)\|}_{\rm F}^2 \mathrm{d}\omega\right)}^{1/2},
\end{align*}
respectively, see, e.\,g., \cite{ZhoDG96} for  details. In many MOR methods one can  bound or even optimize the transfer function error in these norms. 

The \emph{iterative rational Krylov algorithm} (IRKA) aims at minimizing the $\htwo$ error, see \cite{Gugercin2009}. A structure-preserving variant known as IRKA-PH, was derived in~\cite{Gugercin2012}, which is extended to DAEs in~\cite{BeaGM21} using the staircase form in Lemma~\ref{lem:staircase}. The strictly proper part and the Markov parameters of the original transfer functions are identified, such that the standard version of IRKA-PH can be applied. These algorithms are based on the tangential interpolation of the transfer function at iteratively defined interpolation points and tangential directions and the construction of projection spaces using an orthonormalization procedure based on rational Krylov subspaces, see \cite{Antoulas2020} for an exhaustive discussion of such interpolation methods.

Passivity-preserving model reduction methods tailored to DAEs were proposed in~\cite{BeaGM21,Reis2010}.
Positive-real balanced truncation (PRBT) \cite{Reis2010} is based on the computation of the minimal solutions of two \emph{algebraic Riccati equations} (or \emph{Lur'e equations}). These minimal solutions take the role of the Gramians in classical balanced truncation (BT) model reduction, and the balancing and truncation procedure can be performed as in BT. While this method also admits an a priori gap metric error bound \cite{GuiO13}, it generally does not preserve the pH form.

	\section{Our method}%
	\label{sec:our_method}		
        Our method is based on optimizing the parameters (i.\,e., the matrix elements) of a low-order pH-DAE such that its transfer function matches the transfer function of a given large-scale pH-DAE\@ a best as possible. As error measures we consider both the $\htwo$ and the $\hinf$ norms. For this we need to
\begin{enumerate}[(i)]
  \item determine a parameterization that encompasses all the pH-DAE types that are described in Lemma~\ref{lem:staircase},
  \item impose a matching of the polynomial part of the given transfer function to keep the errors well-defined and bounded, and
  \item define update rules for the parameters such that  $\htwo$ or  $\hinf$ errors are minimized.
\end{enumerate}

We set up a parameterization that allows the strictly proper part of the transfer function to be tuned independently of the polynomial part in Section~\ref{sec:parametrization}. This way, matching of the polynomial part and minimizing the errors can be addressed separately. We present several alternative methods of computing the polynomial part of a potentially large-scale pH-DAE in Section~\ref{sub:ourmethod_polcomp}. Methods of both $\hinf$ and $\htwo$ approximation are then given in Sections~\ref{sub:ourmethod_hinf} and~\ref{sub:ourmethod_htwo}, respectively.

\subsection{Parameterization}\label{sec:parametrization}

Before we set up the parameterization, we shall first recall a few properties of transfer functions of passive systems which are positive real.
A real-rational matrix $H$ is referred to as \emph{positive real}, if
\begin{enumerate}[(i)]
\item $H$ has no poles in the open right complex half-plane $\C^+ := \left\{ s \in \C\; |\; \operatorname{Re}(s) > 0 \right\}$;
\item $H(s) + H(s)^\mathsf{H} \ge 0$ for all $s \in \C^+$.
\end{enumerate}
By \cite[Theorem 2.7.2]{AndV73} these two conditions are equivalent to:
\begin{enumerate}[(i)]
 \item The real-rational matrix $H$ has no poles in $\C^+$. 
 \item All poles of $H$ on the extended imaginary axis ${\ri\R \cup \{-\infty,\infty\}}$ are simple. Moreover, if $\ri\omega_0 \in \ri\R$ is a finite pole of $H$, then the residue matrix ${\lim_{s \to \ri\omega_0} (s-\ri\omega_0)H(s)}$ is symmetric positive semi-definite. Similarly, the residue matrix for the poles at infinity $\lim_{\omega \to \infty} \frac{H(\ri\omega)}{\ri\omega}$ is symmetric and positive semi-definite.
 \item For each $\omega \in \R$ for which $\ri\omega$ is not a pole of $H$, it holds that
\begin{equation*}
  H(\ri \omega) + H(\ri \omega)^\mathsf{H} \ge 0.
\end{equation*}
\end{enumerate}
\begin{lem}\label{lem:tf_split}
  Let a real-rational positive real $m \times m$ matrix $H$ be given. We can decompose it into the sum 
  \begin{equation*}
    \pHtf(s) = \pHtf_{\rm sp}(s) + \pol_0 + \pol_1\cdot s,
  \end{equation*}
  where $\pHtf_{\rm sp}$ is \emph{strictly proper} (i.\,e., $\lim_{s\rightarrow\infty}\pHtf_{\rm sp}(s)=0$), $\pol_0,\,\pol_1\in\R^{\enu\times \enu}$, and where
    \begin{enumerate}[(a)]
      \item $\pol_1$ is symmetric positive semidefinite, and
      \item the \emph{proper} part $\pHtfp(s) := \pHtf_{\rm sp}(s) + \pol_0$ is positive real.
    \end{enumerate}
\end{lem}
\begin{proof}
  We refer to~\cite{Wohlers1969} for the proof of (a).
 To show (b), we decompose $\pHtf$ into its proper and improper parts, and for each $\omega \in \R$ for which $\ri\omega$ is not a pole of $H$ we obtain
  \begin{align*}
    \pHtf(\mathsf{i}\omega) + \pHtf{(\mathsf{i}\omega)}^{\mathsf{\pHtf}} & = \pHtfp(\mathsf{i}\omega) + \pHtfp{(\mathsf{i}\omega)}^{\mathsf{\pHtf}} + \mathsf{i}\omega(\pol_1-\pol_1^\T) \\ & = \pHtfp(\mathsf{i}\omega) + \pHtfp{(\mathsf{i}\omega)}^{\mathsf{\pHtf}} \geq 0,
  \end{align*}
  where we have used (a). Consequently, the proper part $\pHtfp$ is positive real. 
\end{proof}
\begin{rem}
 If all finite eigenvalues of the pencil ${sE-(J-R)}$ have negative real part as imposed by our assumptions, then its transfer function $H$ cannot have finite poles on the imaginary axis, in particular, we have $\pHtfp \in \rhinf^{m \times m}$.
\end{rem}

This result allows us to consider the improper part ${\pol_1 \cdot s}$ separately from the proper part of the given transfer function, as the proper part is still positive real and can thus be parameterized separately. For pH-DAE systems as in~\eqref{eq:FOM}, we can even make further assertions about the structure of the proper part $\pHtfp$.
\begin{thm}\label{thm:H_p_is_pH}
  The proper part $\pHtfp$ of the transfer function of any pH-DAE can be realized by an implicit ODE system of the form
  \begin{equation}\label{eq:FOM_prop}
    \begin{aligned}
      \prE \prdx(t) &= (\prJ-\prR)\prx(t) +(\prG-\prP)u(t), \\
      \pry(t) &= {(\prG+\prP)}^\T \prx(t) +(\prS-\prN)u(t),
    \end{aligned}
  \end{equation}
  with $\prx:\R \to \R^{\nx_2}$ and $\prE>0$, where $n_2$ is the corresponding dimension as in Lemma~\ref{lem:staircase}.
\end{thm}
\begin{proof}
  The claim that the proper part of any transfer function of a pH-DAE admits a realization of the form \eqref{eq:FOM_prop} follows immediately, as any proper positive real real-rational matrix is realizable by a passive ODE system, which can in turn be transformed into pH form \cite{BeaMX22}.
  In the Appendix, we show that we can build a realization with state-space dimension $n_2$ by deriving it directly from the staircase form in Lemma~\ref{lem:staircase}. 
\end{proof}

In Theorem~\ref{thm:param} we exploit the presented properties to derive a novel approach to parameterizing  pH-DAEs of differentiation index two. Therein, the functions $\vtu(\cdot)$ and $\vtsu(\cdot)$ map vectors row-wise to appropriately sized upper triangular and strictly upper triangular matrices, respectively. The function names are abbreviations for \emph{vector-to-upper} and \emph{vector-to-strictly-upper}, respectively. The function $\vtf_m(\cdot)$ is a simple reshaping operation, which maps a vector in $\R^{n\cdot m}$ to a matrix in $\R^{n\times m}$. A detailed description of these functions can be found in~\cite[Definition 3.1]{Schwerdtner2020}.

\begin{thm}\label{thm:param}
  Let $\theta \in \R^{\nx_\theta}$ be a parameter vector partitioned as $\theta = \begin{bmatrix}
    \theta_J^\T, \theta_W^\T, \theta_G^\T, \theta_N^\T, \theta_L^\T
  \end{bmatrix}^\T$ 
  with
  $\theta_J \in \R^{\dimx(\dimx-1)/2}$,
  $\theta_W \in \R^{(\dimx+\dimu)(\dimx+\dimu+1)/2}$,
  $\theta_G \in \R^{\dimx \cdot \dimu}$,
  $\theta_N \in \R^{\dimu(\dimu-1)/2}$, and
  $\theta_L \in \R^{\dimu\cdot \diml}$.
  Furthermore, define the matrix-valued functions
  \begin{subequations}
    \begin{align}
      \rprJ(\theta) &:= \vtsu{(\theta_J)}^\T - \vtsu(\theta_J),\\
      W(\theta) &:= \vtu(\theta_W)\vtu{(\theta_W)}^\T  \label{eq:Wconstruction},\\
       \rprR(\theta) &:= \begin{bmatrix} I_{\dimx} & 0 \end{bmatrix} W(\theta) \begin{bmatrix} I_{\dimx} & 0 \end{bmatrix}^\T,\\
      \rprP(\theta) &:= \begin{bmatrix} I_{\dimx} & 0 \end{bmatrix} W(\theta) \begin{bmatrix} 0 & I_{\dimu} \end{bmatrix}^\T,\\
      \rS(\theta) &:= \begin{bmatrix} 0 & I_{\dimu} \end{bmatrix} W(\theta) \begin{bmatrix} 0 & I_{\dimu} \end{bmatrix}^\T, \\
      \rprG(\theta) &:= \vtf_m(\theta_G),\\
      L(\theta) &:= \vtf_\diml(\theta_L),\\
      \rN(\theta) &:= \vtsu{(\theta_N)}^\T - \vtsu(\theta_N).
    \end{align}\label{eq:PHParamMatrices}
  \end{subequations}
  Then the parameter-dependent DAE
  \begin{align}
    \label{eq:pHParam}
    \pHsysr(\theta):
    \begin{cases}
      \!\begin{aligned}
        \rE\rdx(t)= &\,(\rJ(\theta)-\rR(\theta))\rx(t) \\ &\quad \quad + (\rG(\theta)-\rP(\theta)) u(t),
      \end{aligned}\\
      \!\begin{aligned}
        \phantom{\rE}\ry(t)= &\,{(\rG(\theta)+\rP(\theta))}^\T \rx(t)\\ &\quad \quad+ (\rS(\theta)-\rN(\theta)) u(t),
      \end{aligned}\\
    \end{cases}
  \end{align}
  with $\rx$ partitioned as  ${\rx(t) = \begin{bmatrix}
    x_1(t)^\T, x_2(t)^\T, x_3(t)^\T
  \end{bmatrix}^\T}$, where ${x_1(t) \in \R^{r}}$, ${x_2(t) \in \R^{\diml}}$, ${x_3(t) \in \R^{\diml}}$ for each $t \in \R$ and
  \begin{align*}
                \rE &= \mat{I_{\dimx} & 0 & 0 \\ 0 & I_{\diml}  & 0 \\ 0 & 0 & 0}, \\
                \rJ(\theta) &= \mat{\rprJ(\theta) & 0 & 0 \\ 0 & 0 & -I_{\diml}  \\ 0 & I_{\diml}  & 0},\, \rR(\theta) = \mat{\rprR(\theta) & 0 & 0 \\ 0 & 0 & 0 \\ 0 & 0 & 0}, \\
                \rG(\theta) &= \mat{\rprG(\theta) \\ 0 \\ L{(\theta)}^\T},\quad \rP(\theta) = \mat{\rprP(\theta) \\ 0 \\ 0},
  \end{align*}	
  satisfies the pH structure conditions. Its transfer function $\pHtfr$ is given by
  \begin{equation*}
    \pHtfr(s,\theta) = \pHtfrp(s,\theta)+ L(\theta)L{(\theta)}^\T \cdot s,
  \end{equation*}
  where $\pHtfrp(s,\theta)$ denotes the transfer function of the proper pH-ODE system
  \begin{align}
    \label{eq:pHParam_p}
    \pHsysrp(\theta):
    \begin{cases}
      \!\begin{aligned}
        \phantom{_r}\dot{x}_1(t)=\, &(\rprJ(\theta)-\rprR(\theta))x_1(t) \\ &\quad \quad + (\rprG(\theta)-\rprP(\theta)) u(t),
      \end{aligned}\\
      \!\begin{aligned}
        \rpry(t)=\, &{(\rprG(\theta)+\rprP(\theta))}^\T x_1(t)\\ &\quad \quad+ (\rS(\theta)-\rN(\theta)) u(t),
      \end{aligned}\\
    \end{cases}
  \end{align}
\end{thm}
\begin{proof}
  For ease of notation, we omit the argument $\theta$ in the system matrices of $\eqref{eq:PHParamMatrices}$. The fact that~\eqref{eq:pHParam} is a pH-DAE system of index two follows directly from the composition of the reduced-order matrices and the parameterization in~\eqref{eq:PHParamMatrices}. The system may be reformulated as
  \begin{align*}
    \dot{x}_1(t) &= (\rprJ-\rprR)x_1(t) + (\rprG-\rprP)u(t), \\
    \rpry(t) &= (\rprG+\rprP)^\T x_1(t) \\
             & \quad \quad + (\rS-\rN)u(t) + LL^\T \dot{u}(t),
  \end{align*}
  where we have used the fact that $x_3(t)=-\dot{x}_2(t)=L^\T\dot{u}(t)$ (assuming that $u$ is differentiable). 
  This leads to the transfer function
  \begin{equation*}
    \pHtfr(s) = \pHtfrp(s) + LL^\T\cdot s,
  \end{equation*}
  with the proper part
  \begin{multline*}
    \pHtfrp(s)=\\ (\rprG+\rprP)^\T (sI_r-(\rprJ-\rprR))^{-1} (\rprG-\rprP)
                \quad \\ + (\rS-\rN).
  \end{multline*}
  The assertion follows from the fact that $\pHtfrp$ is the transfer function of the system in~\eqref{eq:pHParam_p}, which clearly fulfills the pH structure conditions.
\end{proof}

\begin{rem}
  We highlight that this parameterization naturally carries over to proper pH-DAEs, i.\,e., with either index $\nu < 2$ or $\pG_4=0$. For these systems we have that ${\diml = \operatorname{rank}(L) = 0}$ and therefore it is sufficient to parameterize the system with an implicit pH-ODE as in \eqref{eq:pHParam_p}. This case is discussed in more detail in \cite{MosSMV22}.
\end{rem} 
\begin{rem}
	Note that for the proper subsystem of the reduced model $\pHsysrp(\theta)$, we assume that $\rprE(\theta) = I_r$. This is not a restriction, since every pH-ODE system may be transformed into such a form using, for instance, the Cholesky factor of $\rprE(\theta)$.
\end{rem}

\subsection{Computation of $\pol_0$ and $\pol_1$}%
\label{sub:ourmethod_polcomp}
To keep the $\hinf$ error between the FOM and ROM transfer functions bounded, the polynomial coefficient $\pol_1$ of the FOM transfer function must be matched exactly in the ROM\@. To obtain a bounded $\htwo$ error, both $\pol_0$ and $\pol_1$ must be matched exactly. Since our parameterization allows the polynomial coefficients to be assigned and the remaining free parameters to be independently optimized, it remains to compute $\pol_0$ and $\pol_1$ efficiently for the FOM transfer function.

One approach is to determine $\pol_0$ and $\pol_1$ by the method outlined in~\cite{MehS05}, where these coefficients can be read off a block-Hankel matrix constructed using the solutions of two projected discrete-time Lyapunov equations.  
Another approach consists in transforming the FOM to the almost Kronecker canonical form derived in~\cite{Achleitner2021}, see also Lemma~\ref{lem:kron_form} in the Appendix.

Alternative approaches determine $\pol_0,\,\pol_1$ by sampling the original transfer function $\pHtf$; hence these are independent of the specific representation of the FOM\@. Assume that two distinct imaginary sampling points $s_1 = \ri\omega_1$, $s_2 = \ri\omega_2$ with two sufficiently large values $\omega_1\neq\omega_2$ are given. As shown in \cite{Antoulas2020a}, we have that
\begin{align*}
  \pHtf(\ri\omega_1)-\pHtf(\ri\omega_2) &= \pHtf_{\rm sp}(\ri\omega_1)-\pHtf_{\rm sp}(\ri\omega_2) +(\ri\omega_1-\ri\omega_2)\pol_1 \\ &\approx (\ri\omega_1-\ri\omega_2)\pol_1,
\end{align*}
which follows from $\lim_{s\to\infty}\pHtf_{\rm sp}(s)=0$. We can then obtain an estimate for $\pol_1$ as
\begin{equation*}
  \hat{\pol}_1 = \Real\left(\frac{\pHtf(\ri\omega_1)-\pHtf(\ri\omega_2)}{\ri\omega_1-\ri\omega_2}\right).
\end{equation*}
Similarly, we have that
\begin{align*}
  \frac{\ri\omega_1\pHtf(\ri\omega_1)-\ri\omega_2\pHtf(\ri\omega_2)}{\ri\omega_1-\ri\omega_2} & \approx \pol_0+\ri(\omega_1+\omega_2)\pol_1,
\end{align*}
which yields an estimate for $\pol_0$ given by 
\begin{equation*}
  \hat{\pol}_0 = 	\Real\left(\frac{\ri\omega_1\pHtf(\ri\omega_1)-\ri\omega_2\pHtf(\ri\omega_2)}{\ri\omega_1-\ri\omega_2}\right).
\end{equation*}
The estimation quality generally depends on the choice of $\omega_1,\,\omega_2$, and we can adapt these sampling points in an iterative manner as in \cite{Schwerdtner2020a}, until a certain tolerance is met. Another way of enhancing the accuracy is to include more than two sampling points, using the Loewner framework~\cite{Antoulas2020a}.

\subsection{$\hinf$ approximation}%
\label{sub:ourmethod_hinf}
Since we assume that the pencil $sE-(J-R)$ is regular and its finite eigenvalues are in the open left half plane, we can proceed as in~\cite{Schwerdtner2020} and obtain a good $\hinf$ approximation by minimizing for decreasing values of $\gamma > 0$ the function
\begin{multline}
  \loss(\theta;\pHtf,\gamma,\mathcal{S}) \\ := 
  \frac{1}{\gamma}\sum_{s_i\in \mathcal{S}}
  \sum_{j=1}^{m}{\left({\left[\sigma_j \left(\pHtf(s_i)-\pHtfr(s_i,\theta)\right)-\gamma\right]}_+\right)}^2
  \label{eq:loss}
\end{multline}
with respect to $\theta$, where 
\begin{align*}
  {[ \cdot ]}_+:  \R \rightarrow [0,\infty), \quad x \mapsto 
  \begin{cases}
    x & \text{if } x\ge 0,\\
    0 & \text{if } x<0.
  \end{cases}
\end{align*}
Here, $\mathcal{S} \subset \ri \R$ is a set of sample points at which the original and reduced transfer functions are evaluated and $\sigma_j(\cdot)$ denotes the $j$-th singular value of its matrix argument. 

The procedure for the $\hinf$ approximation pH-DAEs of index two is described in Algorithm~\ref{alg:sobmor}. It is based on repetitively minimizing $\loss$ in conjunction with a bisection over~$\gamma$. Note that $\loss(\cdot;\pHtf,\gamma,\mathcal{S})$ attains its global minimum at zero when the error $\|\pHtf(s_i) - \pHtfr(s_i)\|_2$ at all sample points $s_i \in \mathcal{S}$ is smaller than  $\gamma$. 
With appropriately chosen sample points (an adaptive sampling procedure is proposed in~\cite{Schwerdtner2021}), this in turn is an indication of $\|\pHtf - \pHtfr( \cdot, \theta)\|_{\hinf} < \gamma$. Thus, finding the minimal $\gamma$ such that $\loss(\cdot;\pHtf,\gamma,\mathcal{S})$ can be minimized to zero (as in Algorithm~\ref{alg:sobmor}) is a reasonable strategy for determining a ROM with a small $\hinf$ error. In~\cite{Schwerdtner2020}, we discuss the further benefits of this general approach, also in comparison to a direct minimization of the $\hinf$ norm.

\begin{algorithm}[tbh]
  \LinesNumbered
  \SetAlgoLined
  \DontPrintSemicolon
  \SetKwInOut{Input}{Input}\SetKwInOut{Output}{Output}
  \Input{FOM transfer function $\pHtf$,
  initial ROM transfer function $\pHtfr( \cdot, \theta_0)$ with parameter $\theta_0 \in \R^{n_\theta}$,
  initial sample point set $\mathcal{S} \subset \mathrm{i}\R$,
  upper bound $\gamma_{\rm u} > 0$,
  bisection tolerance $\varepsilon_1 > 0$,
  termination tolerance $\varepsilon_2 > 0$.
}
  \Output{Reduced order model as in Theorem \ref{thm:param} with parameter $\theta_{\rm fin}$.}
  Set $j:=0$ and $\gamma_{\rm l}:=0$.\;
  Compute $\pol_1$ using either method in Section~\ref{sub:ourmethod_polcomp}. \;
  \While{$(\gamma_{\rm u}-\gamma_{\rm l})/(\gamma_{\rm u}+\gamma_{\rm l}) > \varepsilon_1$}{
    Set $\gamma=(\gamma_{\rm u}+\gamma_{\rm l})/2$.\;
    Update sample point set $\mathcal{S}$ using~\cite[Alg.~3.1]{Schwerdtner2021}. \;
    Solve the minimization problem 
    \begin{align*}
      \label{eq:sobmorminprob}
    \alpha &:= \min\limits_{\theta\in \R^{n_\theta}}\loss(\theta;\pHtf,\gamma,\mathcal{S}) \\ & \text{s.\,t.\,} \vtf_\diml{(\theta_L)}^\T \vtf_\diml(\theta_L) = \pol_1
    \end{align*}
    with minimizer $\theta_{j+1} \in \R^{n_\theta}$, initialized at $\theta_j$. \;
  \eIf{$\alpha > \varepsilon_2$}{
    Set $\gamma_{\rm l}:=\gamma$.\;
    }{
    Set $\gamma_{\rm u}:=\gamma$.\;
  }
  Set $j:=j+1$.
  }
  Set $\theta_{\rm fin} := \theta_j$. \;
  Construct the ROM with $\theta_{\rm fin}$ as in Theorem~\ref{thm:param}.\;
  \caption{SOBMOR-$\hinf$}\label{alg:sobmor}
\end{algorithm}

\subsection{$\htwo$ approximation}%
\label{sub:ourmethod_htwo}
For $\mathcal{H}_2$-optimal model reduction, we first have to ensure that ${\pHtf-\pHtfr(\cdot,\theta)\in \mathcal{RH}_{2}^{\enu \times \enu}}$ (see Section \ref{sub:ourmethod_polcomp}). In addition to the condition that $L(\theta)L(\theta)^\T = \pol_1$, we therefore also require that
\begin{align*}
	\rS(\theta) &= \frac{1}{2}(\pol_0^\T+\pol_0) = \pS_{\rm p}, \\
	\rN(\theta) &= \frac{1}{2}(\pol_0^\T-\pol_0) = \pN_{\rm p}.
\end{align*}
This may be achieved by fixing $\theta_L,\,\theta_N$ and $m(m+1)/2$ parameters in $\theta_W$ once $\pol_0$ and $\pol_1$ have been obtained using any of the methods in Section~\ref{sub:ourmethod_polcomp}  (see \cite{MosSMV22} for details). The remaining $n_{\theta}-m(m+\ell)$ parameters in $\theta$ can be tuned to minimize the error $\Vert \pHtf-\pHtfr(\cdot,\theta) \Vert_{\cH_2}$, as initially proposed in \cite{Moser2020}.

Assume that the eigenvalues of $\rprJ(\theta)-\rprR(\theta)$ are simple, and consider the spectral decomposition
\begin{equation} \label{eq:red_evp}
	(\rprJ(\theta)-\rprR(\theta)) Z(\theta) = Z(\theta)\Lambda(\theta),
\end{equation}
where $\Lambda(\theta) = \text{diag}(\reig_1(\theta),\,\ldots\,,\reig_r(\theta))$ and $Z(\theta)$ contains the right eigenvectors as columns. The transfer function $\pHtfr(\cdot,\theta)$ may then be represented by the partial fraction expansion 
\begin{equation*}
	\pHtfr(s,\theta) = \sum_{i=1}^{r} \frac{\rl_i(\theta)\rr_i(\theta)^\T}{s-\reig_i(\theta)}+\rS(\theta)-\rN(\theta)+L(\theta)L(\theta)^\T\cdot s,	
\end{equation*}
 where ${\rl_i(\theta),\,\rr_i(\theta)\in \C^m}$ with
\begin{align*}
	\rl_i(\theta) &= (\rprG(\theta)+\rprP(\theta))^\T Z(\theta) e_i, \\
	\rr_i(\theta) &= (\rprG(\theta)-\rprP(\theta))^\T Z(\theta)^{-\T}e_i,
\end{align*}
and where $e_i$ denotes the $i$-th standard basis vector of $\R^r$. As shown in \cite[Theorem 2.1]{Beattie2009a}, we may then express the $\cH_{2}$ error by	
\begin{multline*}
		{\Vert \pHtf-\pHtfr(\cdot,\theta) \Vert}_{\cH_2}^2  = \\ {\Vert \pHtf_{\rm sp} \Vert}_{\cH_2}^2 
		 - 2\sum_{i=1}^{r}\rl_i(\theta)^\T \pHtf_{\rm sp}(-\reig_i(\theta))\rr_i(\theta) \\ 
		 + \sum_{j,k=1}^{r} \frac{\rl_j(\theta)^\T\rl_k(\theta)\rr_k(\theta)^\T\rr_j(\theta)}{-\reig_j(\theta)-\reig_k(\theta)},
\end{multline*}
where $\pHtf_{\rm sp}$ again denotes the strictly proper part of $\pHtf$. Since ${\Vert \pHtf_{\rm sp} \Vert}_{\cH_2}^2$ does not depend on $\theta$, it can be neglected in the optimization. Consequently, we define the objective functional
\begin{equation*}
	\losst(\theta;\pHtf) :={\Vert \pHtf-\pHtfr(\cdot,\theta) \Vert}_{\cH_2}^2 - {\Vert \pHtf_{\rm sp} \Vert}_{\cH_2}^2
\end{equation*}
which can be evaluated efficiently, since it only requires the solution of the reduced-order eigenvalue problem in \eqref{eq:red_evp} and $r$ evaluations of $\pHtf_{\rm sp}$ at $-\reig_i(\theta)$, $i=1,\,\ldots,\,r$. Note that to evaluate $\pHtf_{\rm sp}$, we do not require a state-space representation of the strictly proper part. Instead, we \emph{indirectly} evaluate $\pHtf_{\rm sp}$ by
\begin{equation}
	\pHtf_{\rm sp}(s) = \pHtf(s) - \pol_0 - \pol_1\cdot s
\end{equation}
for all $s\in\C$. Since the original matrices are sparse, this is much more efficient than an explicit computation of the strictly proper part of the FOM, as shown, for example, in \cite{Achleitner2021}. The objective functional $\losst(\cdot;\pHtf)$ can then be optimized with respect to $\theta$ using gradient-based approaches. For the derivation of the gradient and possible initialization strategies, we refer 
to \cite{MosSMV22} and summarize the procedure in Algorithm \ref{alg:H2OPT}.

\begin{algorithm}[tbh]
	\LinesNumbered
	\SetAlgoLined
	\DontPrintSemicolon
	\SetKwInOut{Input}{Input}\SetKwInOut{Output}{Output}
	\Input{FOM transfer function $\pHtf$, reduced order $r\in\mathbb{N}$.
	}
	\Output{Reduced-order model as in Theorem \ref{thm:param} with parameter $\theta_{\rm fin}$.}
	Compute $\pol_0,\,\pol_1$ (see Section~\ref{sub:ourmethod_polcomp}). \;
	Initialize $\theta_0$ s.\,t. $\rS(\theta_0)-\rN(\theta_0)=\pol_0$ and $L(\theta_0)L(\theta_0)^\T = \pol_1$.\;
	Solve 
	\begin{align*}
		\theta_{\rm fin} &= \argmin\limits_{\substack{\theta\in \R^{n_\theta}}} \losst(\theta; \pHtf) \\
		&\text{s.\,t. } \rS(\theta)-\rN(\theta)=\pol_0,\, L(\theta)L(\theta)^\T = \pol_1.
	\end{align*}\;
	Construct the ROM with $\theta_{\rm fin}$ as in Theorem \ref{thm:param}.\;
	\caption{PROPT-$\mathcal{H}_2$}
	\label{alg:H2OPT}
\end{algorithm}

	\section{Numerical examples}%
	\label{sec:numerical_examples}
        In this section, we illustrate the properties of our new methods using several numerical tests based on well-known benchmarks.

\subsection{Benchmark models}
The first test case we consider is the instationary flow of incompressible fluids on the spatial domain ${\Omega=(0,1)^2}$ with the boundary $\partial \Omega$ and time interval $[0,T]$, as in \cite{Hauschild2019}. The flow is modelled by the Oseen equations
\begin{alignat*}{2}
  \partial_t v &= -(a\cdot\nabla)v+\nu\Delta v - \nabla p+f, \qquad && \text{in } \Omega \times (0,T],\\
  0 &= \dv v, &&\text{in } \Omega \times (0,T],
\end{alignat*}
with velocity vector $v$, pressure $p$, viscosity $\nu>0$, external forces $f$ and a convective term with driving velocity $a$.
As shown in \cite{Hauschild2019}, spatial semi-discretization of the Oseen equations by a finite difference method with associated non-slip boundary conditions and initial velocity $v_0\in\R^2$, i.\,e.,
\begin{alignat*}{2}
  v &= 0,\qquad&& \text{on } \partial\Omega \times (0,T],\\
  v &= v_0,&& \text{in } \Omega \times {0},
\end{alignat*}
leads to a pH-DAE of index two and with $n_3=0$ in the staircase form of Lemma \ref{lem:staircase}. Additionally, we have that $G_4=0$, which makes the transfer function $\pHtf$ strictly proper. We consider two models: \texttt{Oseen-1} with $n=279$ and \texttt{Oseen-2} with ${n=7399}$. 

\begin{figure}[tpb]
  \centering
  \includegraphics[width=0.45\textwidth]{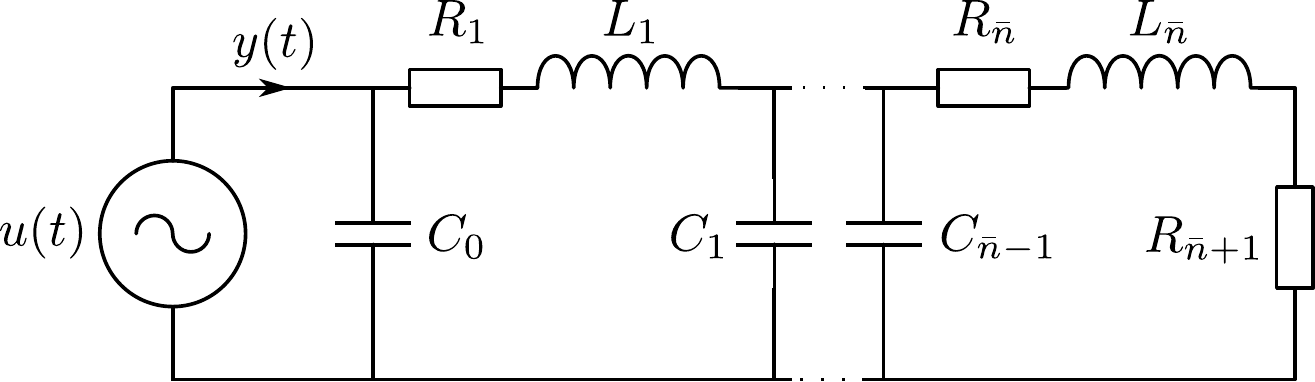}
  \caption{RCL ladder network}\label{fig:RCL_Ladder}
\end{figure}
\begin{figure}[t]
  \centering
  \input{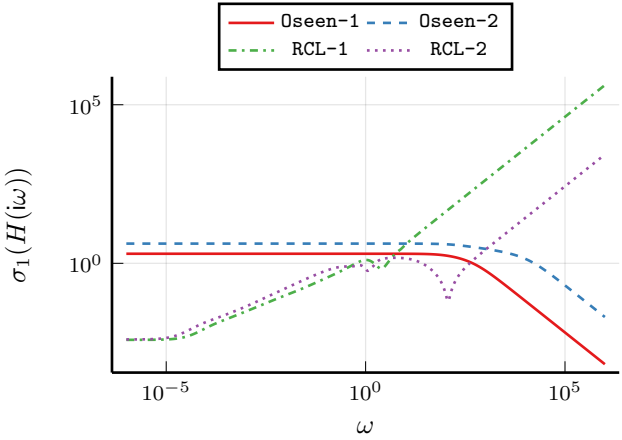} \\
  \caption{Maximum singular values of considered FOM transfer functions}\label{fig:FOMs}
\end{figure}

PH-DAEs of index two also occur in the network modeling of RCL circuits by directed graphs, as described in~\cite{Freund2011}. The second test case we consider relates to RCL ladder networks as shown in Figure~\ref{fig:RCL_Ladder}, which are part of the software package \textsf{PortHamiltonianBenchmarkSystems}\footnote{\url{https://algopaul.github.io/PortHamiltonianBenchmarkSystems/RclCircuits/}}. Here, $\bar{n}$ denotes the number of loops in the system. If we choose the supplied voltage $u$ as the input and the current $y$ as the output, we directly obtain a pH-DAE model as in \eqref{eq:FOM}. The algebraic equation of the model reflects Kirchhoff's voltage law and, due to the loop of the voltage source and the capacitor $C_0$, the transfer function is improper, i.\,e., $\pol_1 \neq 0$. We consider the two examples \texttt{RCL-1} and \texttt{RCL-2} with $\bar{n}=500$, which leads to a model dimension of $n=1502$. Note that in this case, we have that $n_3\neq 0$. The model parameters for the resistances, inductances and capacitances are chosen randomly between $0$ and $1$, and the sigma plots of the resulting FOMs\footnote{All FOM system matrices are available at \url{https://doi.org/10.5281/zenodo.6602125}} are shown in Figure~\ref{fig:FOMs}.

\subsection{Results}

\begin{table}[t]
	\centering
	\caption{Dimensions of benchmark systems}\label{tab:systems}
	\begin{tabular}{c|cccc}
		Model name & $n_1 = n_4$ & $n_2$     & $n_3$ & $n$\\ \hline
		\texttt{Oseen-1}    & $99$    & $81$    & $0$   & $279$ \\
		\texttt{Oseen-2}    & $2499$ & $2401$ & $0$   & $7399$\\
		\texttt{RCL-1}, \texttt{RCL-2}    & $1$    & $999$  & $501$   & $1502$\\
	\end{tabular}
\end{table}

Figure~\ref{fig:NormResults} shows the $\hinf$ and $\htwo$ errors for the \texttt{Oseen-1} model and both RCL systems. To compare our methods with the IRKA-PH and PRBT algorithms, we plot these errors over the dimension $r$ of the \emph{proper} part of the ROM. We highlight that the ROMs of our methods are always minimal realizations with state-space dimension $r+2\diml$, see Theorem \ref{thm:param}. In the case of the PRBT algorithm, minimal realizations can also be computed but may require the solution of discrete-time projected Lyapunov equations, see \cite{Reis2010}. Also note that the ROMs produced by PRBT are generally not in pH form. 

\begin{figure*}[tb]
  \centering
  \begin{tabular}{cc}
    \begin{tikzpicture}[/tikz/background rectangle/.style={fill={rgb,1:red,1.0;green,1.0;blue,1.0}, draw opacity={1.0}}, show background rectangle]
\begin{axis}[point meta max={nan}, point meta min={nan}, legend cell align={left}, legend columns={2}, title, title style={at={{(0.5,1)}}, anchor={south}, font={{\fontsize{14 pt}{18.2 pt}\selectfont}}, color={rgb,1:red,0.0;green,0.0;blue,0.0}, draw opacity={1.0}, rotate={0.0}}, legend style={color={rgb,1:red,0.0;green,0.0;blue,0.0}, draw opacity={1.0}, line width={1}, solid, fill={rgb,1:red,1.0;green,1.0;blue,1.0}, fill opacity={1.0}, text opacity={1.0}, font={{\fontsize{8 pt}{10.4 pt}\selectfont}}, text={rgb,1:red,0.0;green,0.0;blue,0.0}, cells={anchor={center}}, at={(0.5, 1.02)}, anchor={south}}, axis background/.style={fill={rgb,1:red,1.0;green,1.0;blue,1.0}, opacity={1.0}}, anchor={north west}, xshift={1.0mm}, yshift={-1.0mm}, clip mode={individual}, width={0.45\textwidth}, height={0.3\textwidth}, scaled x ticks={false}, xlabel={Reduced model order $r$}, x tick style={color={rgb,1:red,0.0;green,0.0;blue,0.0}, opacity={1.0}}, x tick label style={color={rgb,1:red,0.0;green,0.0;blue,0.0}, opacity={1.0}, rotate={0}}, xlabel style={}, xmajorgrids={true}, xmin={0.7537999999999998}, xmax={10.18620000000001}, xtick={{1.0,2.0,3.0,4.0,5.0,6.0,7.0,8.0,9.0,10.0,11.77,14.54000000000001,16.54000000000001,18.54000000000001,20.54000000000001,22.54000000000001}}, xticklabels={{1,2,3,4,5,6,7,8,9,10,$\cdots$,200,202,204,206,208}}, xtick align={inside}, xticklabel style={font={{\fontsize{8 pt}{10.4 pt}\selectfont}}, color={rgb,1:red,0.0;green,0.0;blue,0.0}, draw opacity={1.0}, rotate={0.0}}, x grid style={color={rgb,1:red,0.0;green,0.0;blue,0.0}, draw opacity={0.1}, line width={0.5}, solid}, axis x line*={left}, x axis line style={color={rgb,1:red,0.0;green,0.0;blue,0.0}, draw opacity={1.0}, line width={1}, solid}, scaled y ticks={false}, ylabel={\scriptsize{$\mathcal{H}_\infty$ error}}, y tick style={color={rgb,1:red,0.0;green,0.0;blue,0.0}, opacity={1.0}}, y tick label style={color={rgb,1:red,0.0;green,0.0;blue,0.0}, opacity={1.0}, rotate={0}}, ylabel style={}, ymode={log}, log basis y={10}, ymajorgrids={true}, ymin={9.608680704523704e-12}, ymax={13.149811833137049}, ytick={{1.0,0.01,0.0001,1.0e-6,1.0e-8,1.0e-10}}, yticklabels={{$10^{0}$,$10^{-2}$,$10^{-4}$,$10^{-6}$,$10^{-8}$,$10^{-10}$}}, ytick align={inside}, yticklabel style={font={{\fontsize{8 pt}{10.4 pt}\selectfont}}, color={rgb,1:red,0.0;green,0.0;blue,0.0}, draw opacity={1.0}, rotate={0.0}}, y grid style={color={rgb,1:red,0.0;green,0.0;blue,0.0}, draw opacity={0.1}, line width={0.5}, solid}, axis y line*={left}, y axis line style={color={rgb,1:red,0.0;green,0.0;blue,0.0}, draw opacity={1.0}, line width={1}, solid}, colorbar={false}]
    \addplot[color={rgb,1:red,1.0;green,0.498;blue,0.0}, name path={80402eac-331b-4a02-9733-a2069cd2ff7c}, draw opacity={1.0}, line width={1}, solid, mark={triangle*}, mark size={3.0 pt}, mark repeat={1}, mark options={color={rgb,1:red,0.0;green,0.0;blue,0.0}, draw opacity={1.0}, fill={rgb,1:red,1.0;green,0.498;blue,0.0}, fill opacity={1.0}, line width={0.75}, rotate={180}, solid}]
        table[row sep={\\}]
        {
            \\
            1.0  0.18449012115835947  \\
            2.0  0.015134734359492197  \\
            3.0  0.0030589893113097677  \\
            4.0  0.0005204865261038832  \\
            5.0  0.00014043233290292762  \\
            6.0  4.632615663325906e-5  \\
            7.0  3.5645766356769794e-6  \\
            8.0  1.9118811335834095e-6  \\
            9.0  5.332962913391085e-7  \\
            10.0  5.5819860122824014e-8  \\
        }
        ;
    \addlegendentry {IRKA-PH$ $}
    \addplot[color={rgb,1:red,0.2157;green,0.4941;blue,0.7216}, name path={8ddc2644-c79b-4130-ae8e-6d221d030e28}, draw opacity={1.0}, line width={1}, solid, mark={diamond*}, mark size={3.0 pt}, mark repeat={1}, mark options={color={rgb,1:red,0.0;green,0.0;blue,0.0}, draw opacity={1.0}, fill={rgb,1:red,0.2157;green,0.4941;blue,0.7216}, fill opacity={1.0}, line width={0.75}, rotate={0}, solid}]
        table[row sep={\\}]
        {
            \\
            1.0  0.18494461389405548  \\
            2.0  0.014041235918563964  \\
            3.0  0.001621575101879713  \\
            4.0  0.00018765565074379056  \\
            5.0  1.4802112946862595e-5  \\
            6.0  1.3878423928289722e-6  \\
            7.0  8.585543209704838e-8  \\
            8.0  1.0580446164957472e-8  \\
            9.0  7.478158717266759e-10  \\
            10.0  2.685443383507527e-10  \\
        }
        ;
    \addlegendentry {PROPT-$\mathcal{H}_2$}
    \addplot[color={rgb,1:red,0.302;green,0.6863;blue,0.2902}, name path={4a8e676a-a471-4748-87e3-db75f124fcec}, draw opacity={1.0}, line width={1}, solid, mark={*}, mark size={3.0 pt}, mark repeat={1}, mark options={color={rgb,1:red,0.0;green,0.0;blue,0.0}, draw opacity={1.0}, fill={rgb,1:red,0.302;green,0.6863;blue,0.2902}, fill opacity={1.0}, line width={0.75}, rotate={0}, solid}]
        table[row sep={\\}]
        {
            \\
            1.0  0.09962006794632873  \\
            2.0  0.0065767760495581895  \\
            3.0  0.0006819448840999752  \\
            4.0  7.389568280584856e-5  \\
            5.0  6.14973049490532e-6  \\
            6.0  5.747528369836662e-7  \\
            7.0  3.749058828429728e-8  \\
            8.0  1.9534827066878593e-9  \\
            9.0  2.5499239148065005e-10  \\
            10.0  2.0032215715936352e-10  \\
        }
        ;
    \addlegendentry {SOBMOR-$\mathcal{H}_\infty$}
    \addplot[color={rgb,1:red,0.651;green,0.3373;blue,0.1569}, name path={e602a9c1-47c0-4643-b9b7-9d1eb92c5e1d}, draw opacity={1.0}, line width={1}, solid, mark={triangle*}, mark size={3.0 pt}, mark repeat={1}, mark options={color={rgb,1:red,0.0;green,0.0;blue,0.0}, draw opacity={1.0}, fill={rgb,1:red,0.651;green,0.3373;blue,0.1569}, fill opacity={1.0}, line width={0.75}, rotate={0}, solid}]
        table[row sep={\\}]
        {
            \\
            1.0  0.45688156804245555  \\
            2.0  0.04091345864577163  \\
            3.0  0.0035524024506317976  \\
            4.0  0.00043413303175544047  \\
            5.0  3.8769674754293795e-5  \\
            6.0  4.011072238885387e-6  \\
            7.0  2.5622843400360756e-7  \\
            8.0  1.5822204649323407e-8  \\
            9.0  8.918852076109384e-10  \\
            10.0  4.0188878150886556e-11  \\
        }
        ;
    \addlegendentry {PRBT$ $}
\end{axis}
\end{tikzpicture} &
    \input{./PlotSources/H2ErrorComparison-Oseen279.tikz} \\
    (a) $\hinf$ errors for the \texttt{Oseen-1} model & (b) $\htwo$ errors for the \texttt{Oseen-1} model \vspace{0.3cm} \\
    \input{./PlotSources/HinfErrorComparison-Index2RandomParams500.tikz} &
    \input{./PlotSources/HinfErrorComparison-DifficultIndex2RandomParams.tikz} \\
    (c) $\hinf$ errors for the \texttt{RCL-1} model & (d) $\hinf$ errors for the \texttt{RCL-2} model \\
  \end{tabular}
  \caption{$\hinf$ and $\htwo$ error comparison for the \texttt{Oseen-1} model and both RCL benchmark models}\label{fig:NormResults}
\end{figure*}

Let us first consider the MOR results on the strictly proper \texttt{Oseen-1} model, shown in Figures~\ref{fig:NormResults}~(a) and~(b). It can be observed that both PROPT-$\mathcal{H}_2$ and SOBMOR-$\hinf$ lead to more accurate models in both the $\htwo$ and the $\hinf$ norms compared to IRKA-PH, especially when the reduced model order is increasing. Compared with PRBT, SOBMOR-$\hinf$ leads to slightly smaller $\hinf$ errors except for $r=10$, while PROPT-$\mathcal{H}_2$ leads to slightly smaller $\htwo$ errors up to reduced orders of $8$.

Figures~\ref{fig:NormResults}~(c) and~(d) show the results for the RCL models. While the difference between the improper FOM and improper ROM transfer function is itself proper, since we ensure a matching of the polynomial parts, the computation of the $\htwo$ norm using the Matlab function \texttt{h2norm} fails. Therefore, we only report the $\hinf$ errors. In terms of the $\hinf$ norm, the ROMs computed with SOBMOR-$\hinf$ have the smallest errors for both RCL models and all reduced model orders. The $\hinf$ error of the PROPT-$\mathcal{H}_2$ ROMs stagnates after a reduced model order of 16 and is worse in general. However, this is expected, since PROPT-$\mathcal{H}_2$ aims at minimizing the $\htwo$ error.

\begin{figure}[tb]
  \centering
  \begin{tabular}{c}
    \input{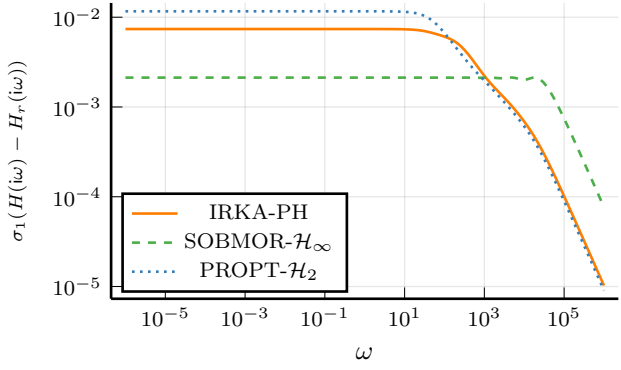} \\
    (a) $r=5$ \vspace{0.3cm}\\
    \input{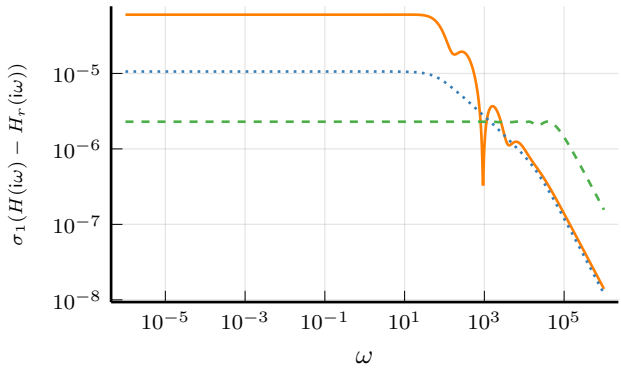} \\
    (b) $r=10$ \\
  \end{tabular}
  \caption{Frequency errors (measured by the maximum singular value) for the \texttt{Oseen-2} model}\label{fig:FreqErrs}
\end{figure}

Figure~\ref{fig:FreqErrs} shows the results for the \texttt{Oseen-2} model. Since no implementation of PRBT that exploits sparsity is currently publicly available, we only report results obtained with our methods and IRKA-PH\@. Our findings are comparable to the smaller \texttt{Oseen-1} model: The error transfer functions of IRKA-PH and PROPT-$\mathcal{H}_2$ are similar for the small reduced model order $r=5$, while for $r=10$, the PROPT-$\mathcal{H}_2$ error is clearly smaller than the error of the IRKA-PH model across a wide frequency range. SOBMOR-$\hinf$ aims at minimizing the maximum error across all frequencies and it can be observed that the maximum errors of the SOBMOR-$\hinf$ ROMs are well below the maximum errors of the ROMs obtained with the other methods for both reduced orders.

	\section{Conclusion}\label{sec:conclusion}
        
        We have presented a flexible MOR approach for $\cH_2$ and $\cH_{\infty}$ approximation of higher index pH-DAEs. Our approach is based on a novel parametrization that can provide a pH realization for any pH descriptor system with an efficient representation of the algebraic part. An adaptation of previously created optimization-based MOR methods allows for the approximation of potentially improper transfer functions. A comparison to state-of-the-art methods shows that our optimization-based approach leads to highly accurate ROMs that are minimal realizations and guaranteed to fulfill the pH structural constraints.
	
	\section*{Acknowledgments}
        {We thank Serkan Gugercin (Virginia Tech) for providing the Matlab code to generate the Oseen models.}

	\bibliographystyle{elsarticle-num} 
	\newpage
	\bibliography{references}
    
    \newpage
		
	\section*{Appendix}
        The following Kronecker-like form was derived in \cite{Achleitner2021}, based on the staircase form. 
\begin{lem}\label{lem:kron_form}
  Consider a regular pH-DAE in staircase form~\eqref{eq:FOM_stair} and define ${\wt{A}:=\wt{J}-\wt{R}}$, ${\wt{B}:=\wt{G}-\wt{P}}$, ${\wt{C}:=(\wt{G}+\wt{P})^\T}$, ${\wt{D}:=\wt{S}-\wt{N}}$. Then there exist nonsingular matrices ${T_1,\,T_2\in\R^{n\times n}}$ such that the pH-DAE may be transformed to a general linear time-invariant system of the form
  \begin{equation} \label{eq:FOM_kron}
    \begin{aligned}
      \check{E}\dot{\check{x}}(t) &= \check{A}\check{x}(t)+\check{B}u(t), \\
      y(t) &= \check{C}\check{x}(t)+\check{D}u(t),
    \end{aligned} 
  \end{equation}		
  where
  \begin{align*}
    \check{E} &:= T_1\wt{E} T_2 = \mat{\check{E}_{11} & 0 & 0 & 0 \\ 0 & \check{E}_{22} & 0 & 0 \\ 0 & 0 & 0 & 0 \\ 0 & 0 & 0 & 0},\\
    \check{A} &:= T_1\wt{A}T_2 = \mat{0 & 0 & 0 & I_{n_1} \\ 0 & \check{A}_{22} & 0 & 0 \\ 0 & 0 & I_{n_3} & 0 \\ -I_{n_4} & 0 & 0 & 0},\\ \check{B} &:= T_1\wt{B} = \mat{\check{B}_1 \\ \check{B}_2 \\ \check{B}_3 \\ \check{B}_4}, \quad \check{C} := \wt{C} T_2 = \mat{\check{C}_1^\T \\ \check{C}_2^\T \\ \check{C}_3^\T \\ \check{C}_4^\T}^\T
  \end{align*}
  and $\check{D} = \wt{D}$. The matrices are partitioned in the same way as in Lemma~\ref{lem:staircase} and, if present, the diagonal block matrices $\check{E}_{11},\check{E}_{22}$ are symmetric positive definite.
\end{lem}
\begin{proof}[Proof of Theorem \ref{thm:H_p_is_pH}]
  Considering the transformed system \eqref{eq:FOM_kron} and a block diagonalization of ${s\check{E}-\check{A}}$ yield the transfer function
  \begin{align*}
    \pHtf(s) &= \check{C}_2(s\check{E}_{22}-\check{A}_{22})^{-1}\check{B}_2 + \pol_0 + \pol_1\cdot s,
  \end{align*}
  where
  \begin{align*}
    \pol_0	&= \check{D}+\check{C}_1\check{B}_4-\check{C}_3\check{B}_3-\check{C}_4\check{B}_1, \\
   	\pol_1 &= \check{C}_4\check{E}_{11}\check{B}_4,
  \end{align*}
  which reveals the split into the proper and improper parts, respectively. From the definition of $T_1,\, T_2$ in \cite{Achleitner2021} we have that
  \begin{align*}
    \check{E}_{11} &= \wt{E}_{11}-\wt{E}_{12}\wt{E}_{22}^{-1}\wt{E}_{21} > 0, \\
    \check{B}_4 &= -\wt{A}_{41}^{-1}\wt{B}_4 = J_{14}^{-\T}G_4,\\	
    \check{C}_4 &= \wt{C}_4	\wt{A}_{14}^{-1} = G_4^\T J_{14}^{-1} = \check{B}_4^\T,
  \end{align*}
  which proves that $\pol_1 = \check{C}_4\check{E}_{11}\check{B}_4 = \pol_1^\T \geq 0$. For ${n_2=0}$, the claim follows immediately from $H(0)+H(0)^\mathsf{H} = \pol_0 + \pol_0^\T \ge 0$. Now let us first consider the most general case where $n_2,n_3>0$. The remaining block matrices from \eqref{eq:FOM_kron} are then given by
  \begin{align*}
    \check{E}_{22} &= \wt{E}_{22}, \\
    \check{A}_{22} &= \wt{A}_{22}-\wt{A}_{23}\wt{A}_{33}^{-1}\wt{A}_{32},\\
    \check{B}_1 &= \wt{B}_1-\wt{A}_{13}\wt{A}_{33}^{-1}\wt{B}_3+(-\wt{A}_{11}+\wt{A}_{13}\wt{A}_{33}^{-1}\wt{A}_{31})\wt{A}_{41}^{-1}\wt{B}_4,\\
    \check{B}_2 &= \wt{B}_2-\wt{A}_{23}\wt{A}_{33}^{-1}\wt{B}_3+(-\wt{A}_{21}+\wt{A}_{23}\wt{A}_{33}^{-1}\wt{A}_{31})\wt{A}_{41}^{-1}\wt{B}_4,\\
    \check{B}_3 &= \wt{A}_{33}^{-1}\wt{B}_3-\wt{A}_{33}^{-1}\wt{A}_{31}\wt{A}_{41}^{-1}\wt{B}_4,\\	
    \check{C}_1 &= \wt{C}_1,\\
    \check{C}_2 &= \wt{C}_2 - \wt{C}_3\wt{A}_{33}^{-1}\wt{A}_{32} +	\wt{C}_4\wt{A}_{14}^{-1}(-\wt{A}_{12}+\wt{A}_{13}\wt{A}_{33}^{-1}\wt{A}_{32}),\\
    \check{C}_3 &= \wt{C}_3,
  \end{align*}
  which again follows from the definition of $T_1,\, T_2$ in \cite{Achleitner2021}. Define the matrices $\wt{\Gamma}, \wt{W} \in\R^{(n+m)\times (n+m)}$ 
  \begin{alignat*}{2}
    \wt{\Gamma} &:= \mat{-\wt{J} & -\wt{G} \\ \wt{G}^\T & -\wt{N}} &&= -\wt{\Gamma}^\T, \\
    \wt{W} &:= \mat{\wt{R} & \wt{P} \\ \wt{P}^\T & \wt{S}} &&= \wt{W}^\T \geq 0,
  \end{alignat*}
  and let the matrices be partitioned as in Lemma \ref{lem:staircase}. Our proof is based on the observation that 
  \begin{equation*}
    \mat{-\wt{A} & -\wt{B} \\ \wt{C} & \wt{D}}  = \mat{-(\wt{J}-\wt{R}) & -(\wt{G}-\wt{P}) \\ (\wt{G}+\wt{P})^\T & \wt{S}-\wt{N}} = \wt{\Gamma} + \wt{W}.
  \end{equation*}
  This is a natural generalization of a similar observation for linear dissipative Hamiltonian systems (see \cite{Guducu2021}) to port-Hamiltonian systems with power-collocated input-output pairs. We will now show that such a decomposition into skew-symmetric and symmetric positive semidefinite parts not only exists for the proper subsystem as well but may be obtained by \emph{structure-preserving} manipulations of the sum $\wt{\Gamma} + \wt{W}$. 

  At first, let $P_{\pi}\in\R^{(n+m)\times(n+m)}$ define a permutation matrix which permutes the third and fifth block rows and columns in the sum $\wt{\Gamma}+\wt{W}$ such that
  \begin{equation*}
    \Psi =  P_{\pi}^\T \big(\wt{\Gamma} + \wt{W}\big) P_{\pi} = \mat{\Psi_{\rm uu} & \Psi_{\rm ul} \\ \Psi_{\rm lu} & \Psi_{\rm ll}},
  \end{equation*} 
  with $\Psi_{\rm ll}=-\wt{J}_{33}+\wt{R}_{33}=-\wt{A}_{33}$. Since the matrix $\wt{A}_{33}$ is nonsingular, we may block-diagonalize $\Psi$ with invertible matrices ${X_1 = \begin{bsmallmatrix} I_n & \Psi_{\rm ul}\Psi_{\rm ll}^{-1} \\ 0 & I_m \end{bsmallmatrix}}$, ${X_2 = \begin{bsmallmatrix} I_n & 0 \\ \Psi_{\rm ll}^{-1} \Psi_{\rm lu} & I_m \end{bsmallmatrix} \in\R^{(n+m)\times(n+m)}}$ such that
  \begin{equation*}
    \Xi = X_1 \Psi X_2 = \mat{\Psi_{\rm uu}-\Psi_{\rm ul}\Psi_{\rm ll}^{-1}\Psi_{\rm lu} & 0 \\ 0 & \Psi_{\rm ll}}.
  \end{equation*}
  Note that $\Xi$ still has a positive semidefinite symmetric part since the Schur complement preserves this property \cite[Corollary 4.3]{Guducu2021}. Finally, it is easy to show that we may compute the proper system matrices via a transformation of $\Xi$ with the full-rank matrix $U\in\R^{(n+m)\times(n_2+m)}$ such that
  \begin{equation*}
    \mat{-\check{A}_{22} & -\check{B}_2 \\ \check{C}_2 & \pol_0} = U^\T \Xi U, \text{ where } U = \mat{ 0 & \check{B}_4 \\ I_{m} & 0 \\ 0 & I_{n_2} \\ 0 & 0\\ 0 & 0}.
  \end{equation*}
 Hence, we obtain the proper system matrices by a series of permutations, block-diagonalization via Schur complements and congruence transformations of $\wt{\Gamma} + \wt{W}$. Since each of these manipulations preserves the positive semidefiniteness of the symmetric part, we obtain a port-Hamiltonian representation of the proper subsystem via
  \begin{align*}
    \mat{-J_{\rm p} & -G_{\rm p} \\ G_{\rm p}^\T & -N_{\rm p}} &= \frac{1}{2}\left(\mat{-\check{A}_{22} & -\check{B}_2 \\ \check{C}_2 & \pol_0} -\mat{-\check{A}_{22} & -\check{B}_2 \\ \check{C}_2 & \pol_0}^\T\right), \\ \mat{R_{\rm p} & P_{\rm p} \\ P_{\rm p}^\T & S_{\rm p}} &= \frac{1}{2}\left(\mat{-\check{A}_{22} & -\check{B}_2 \\ \check{C}_2 & \pol_0} +\mat{-\check{A}_{22} & -\check{B}_2 \\ \check{C}_2 & \pol_0}^\T\right).
  \end{align*}
  The fact that ${E_{\rm p}=\check{E}_{22}}> 0$ proves the claim for ${n_3>0}$. Similar arguments apply to  pH-DAEs of differentiation index two for which $n_3=0$ and where we have that
  \begin{equation*}
    \mat{-\check{A}_{22} & -\check{B}_2 \\ \check{C}_2 & \pol_0} = U^\T (\wt{\Gamma} + \wt{W}) U, \text{ where } U = \mat{ 0 & \check{B}_4 \\ I_{n_2} & 0 \\ 0 & 0\\ 0 & I_{m}}.
  \end{equation*}
This concludes the proof.

\end{proof}

\end{document}